\documentclass[10pt,conference]{IEEEtran}
\usepackage{authblk}
\usepackage{graphicx}
\usepackage{url}
\usepackage{amsmath, amsthm, amssymb}
\usepackage{subcaption}
\usepackage{caption}
\usepackage{algorithm}
\usepackage{multirow}
\usepackage{algpseudocode}
\usepackage[T1]{fontenc}
\usepackage{array}
\usepackage{booktabs}
\usepackage{latexsym}

\usepackage{enumitem}

\algrenewcommand\Return{\State \algorithmicreturn{ }}

\graphicspath{{./figures/}}

\usepackage{color}

\begin{document}

\title{Towards Network-Failure-Tolerant Content Delivery for Web Content}


\author[1]{Wen Hu}
\author[2]{Zhi Wang}
\author[1]{Lifeng Sun}
\affil[1]{Tsinghua National Laboratory for Information Science and Technology \authorcr Department of Computer Science and Technology, Tsinghua University }
\affil[2]{Graduate School at Shenzhen, Tsinghua University}
\affil[ ]{\{hu-w12@mails., wangzhi@sz., sunlf@\}tsinghua.edu.cn}

\maketitle

\begin{abstract}

  Popularly used to distribute a variety of multimedia content items in today's Internet, HTTP-based web content delivery still suffers from various content delivery failures, including server failures~\cite{facebook}, network failures \cite{understand_network_failure} and routing failures~\cite{OSPF}. Hindered by the expensive deployment cost, the conventional CDN can not deploy as many edge servers as possible to successfully deliver content items to all users under these delivery failures. In this paper, we propose a joint CDN and peer-assisted web content delivery framework to address the delivery failure problem. Different from conventional peer-assisted approaches for web content delivery, which mainly focus on alleviating the CDN servers' bandwidth load, we study how to use a browser-based peer-assisted scheme, namely WebRTC, to resolve content delivery failures. To this end, we carry out large-scale measurement studies on how users access and view webpages. Our measurement results demonstrate the challenges (e.g., peers stay on a webpage extremely short) that can not be directly solved by conventional P2P strategies, and some important webpage viewing patterns (e.g., predictability of users' webpage viewing time). Due to these unique characteristics, WebRTC peers open up new possibilities for helping the web content delivery, coming with the problem of how to utilize the dynamic resources efficiently. We formulate the peer selection that is the critical strategy in our framework, as an optimization problem, and design a heuristic algorithm based on the measurement insights to solve it. Our simulation experiments driven by the traces from Tencent QZone, one of the most popular social service platforms in China, demonstrate the effectiveness of our design: compared with non-peer-assisted strategy and random peer selection strategy, our design significantly improves the successful relay ratio of web content items under network failures, e.g., our design improves the content download ratio up to $60\%$ even when users located in a particular region (e.g., city) where none can connect to the regional CDN server.

\end{abstract}


\section{Introduction} \label{sec:intro}

Dominating today's Internet traffic, HTTP-based online services are suffering from many types of content delivery failures (e.g., server failures~\cite{facebook}, network failures \cite{understand_network_failure}, routing failures \cite{OSPF}), significantly degrading the quality-of-service (QoS) provisioned to online users. For example, Google service's outage led to a traffic drop of around $40\%$ during $5$ minutes in $2013$. This problem has been exacerbated when today's online services are designed to utilize aggregated content items and information, which are served by a geo-distributed Content Delivery Network (CDN), e.g., a web page consisting of content items that are served by multiple servers may fail to render at the client side if it fails to download only parts of the content items from some servers.  

Conventional approaches to solve this problem are summarized as follows: (1) Replicating content items to multiple servers at different locations to avoid the failures of the servers \cite{web_replication}, e.g., a user will be directed to download the same content from a different server when the original requested server does not respond. (2) Deploying edge-network proxies \cite{web_cache}, e.g., a user can download content items from the nearby gateway proxy which may cache the requested content items. However, such approaches: (1) require additional replication of content items, causing additional deployment and operation costs; (2) are primarily for resolving the problem of static content items delivery; (3) are infeasible for today's websites, which contain not only static content items, such as images, videos and SWF (Flash objects), but also dynamic content items~\cite{zhi-tmm-cpcdn2014} which include personalized elements---intrinsic un-cacheable feature of dynamic content items degrades the performance of these approaches.

\begin{figure}[t]
       \begin{minipage}[t]{.35\linewidth}
	         \centering
	               \includegraphics[width=\linewidth]{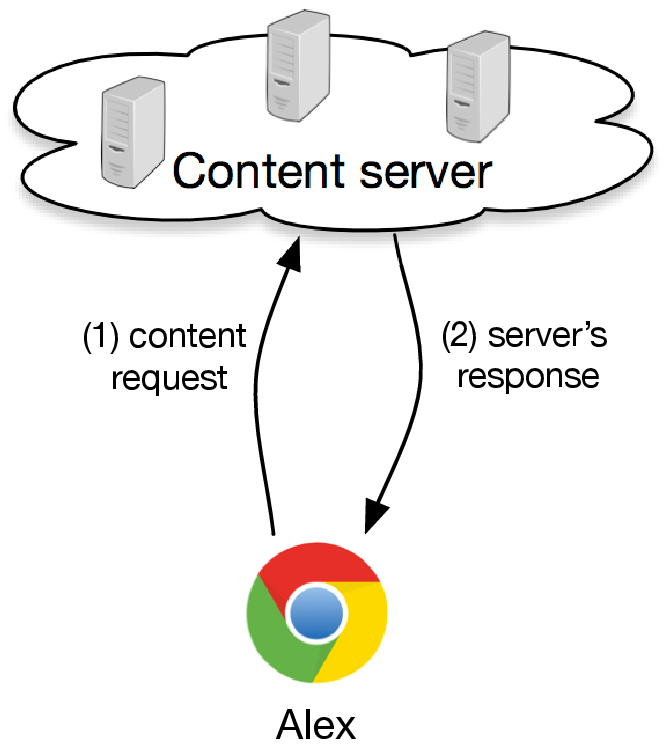}
				   \raggedright
	               \scriptsize (a) Conventional web content delivery.
	       \end{minipage}
         \hfill
       \begin{minipage}[t]{.6\linewidth}
             \centering
            \includegraphics[width=\linewidth]{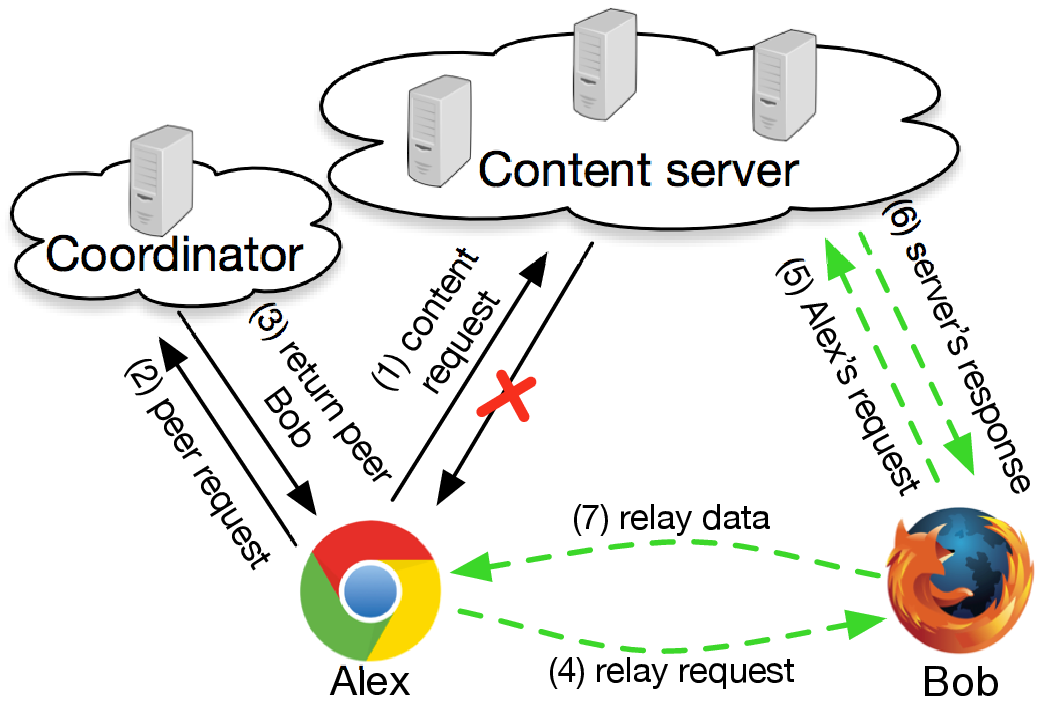}
			\raggedright
                   \scriptsize (b) CDN+WebRTC web content delivery.
        \end{minipage}
       \caption{Comparison between conventional content delivery and CDN+WebRTC content delivery.}
       \label{fig:data_relay_framework}
\end{figure}

In this paper, we propose a joint CDN and peer-assisted web content delivery framework based on WebRTC \cite{webrtc}, which is defined by World Wide Web Consortium (W3C) and the Internet Engineering Task Force (IETF), and has been widely supported by major browser vendors \cite{alexandru2014impact}, e.g., Google, Opera, Mozilla, Microsoft (under development). As illustrated in Fig.~\ref{fig:data_relay_framework}, by strategically selecting appropriate WebRTC-powered browsers to form the ``recovery'' delivery path (denoted as green dotted lines), these widely deployed browsers can assist in content delivery when a user fails to receive requested content items from the original CDN servers. As opposed to the conventional content delivery path, the recovered delivery path is not only from the CDN server through backbone networks to end users, but also from end users to end
users. In order to select the most appropriate peer to recover content delivery failures, we perform exhaustive measurement studies on users' webpage viewing behaviors, and the web-session characteristics.


Our contributions in this paper are summarized as follows:

$\rhd$ We conduct large-scale measurement studies to demonstrate the content delivery failures in today's web content delivery to motivate our design. We also study users' webpage viewing behaviors and the web-session characteristics, including: (1) high churning rate of online peers; (2) low probability of ``re-join'', i.e., re-visit the same webpage later; (3) fragmentation of web content, i.e., various content types (e.g., JavaScript, HTML, CSS, image, Flash, etc) are placed on different servers. These motivate us to propose a novel peer selection strategy customized for our CDN and peer-assisted web content delivery design, but not directly adopt the traditional P2P strategies.

$\rhd$ Based on our measurement studies, we propose a joint CDN and peer-assisted web content delivery framework and mathematically formulate the peer selection problem as an optimization problem. The objective is to choose a set of peering WebRTC users to form a delivery path in case of network failures. A heuristic algorithm is designed to solve this problem.

$\rhd$ We carry out simulation-based experiments to test the effectiveness of our design under dynamic and extreme running scenarios. Our design improves the content download ratio up to $60\%$ even when users located in a region (e.g., a city) where none can connect to the regional CDN server.


The rest of the paper is structured as follows. In Section \ref{section:measurement}, we carry out some measurement studies to demonstrate the motivation and design principles. We present the problem formulation and our design in Section \ref{section:algorithm}. In Section \ref{section:experiment}, we evaluate the performance of our system. Section \ref{section:relatedwork} discusses the related work. We conclude this work in Section \ref{section:conclusion}.

\section{Measurement-driven Motivation and Design Principles}
\label{section:measurement}

In this section, we present our motivation, and the design principles learnt from measurement studies.

\subsection{Measurement Methodology}

We first present how we carry out the measurement studies. We study users' webpage accessing patterns based on real world traces collected by country-wide deployed CDN servers of Tencent QZone~\cite{qzone}. The traces consist of the user visiting information of two testing webpages in Tencent QZone, including $118,707$ webpage visits, in which $2,300$ sessions have encountered network failures. Each trace item contains the following information: (1) the user identifier, (2) the timestamp when a user begins to request the webpage, (3) the timestamp when a user leaves the current webpage, (4) the indicator of content fetching failure (i.e., whether a content item contained in a webpage can not be downloaded from the original server). Based on these traces, we are able to study the challenges in peer-assisted web content delivery and the design principles.


\subsection{Strong Randomness of Web Content Delivery Failures}

\begin{figure*}[t]
     \begin{minipage}[t]{0.62\linewidth}
		 \begin{subfigure}[b]{.49\linewidth}
               \includegraphics[width=\linewidth]{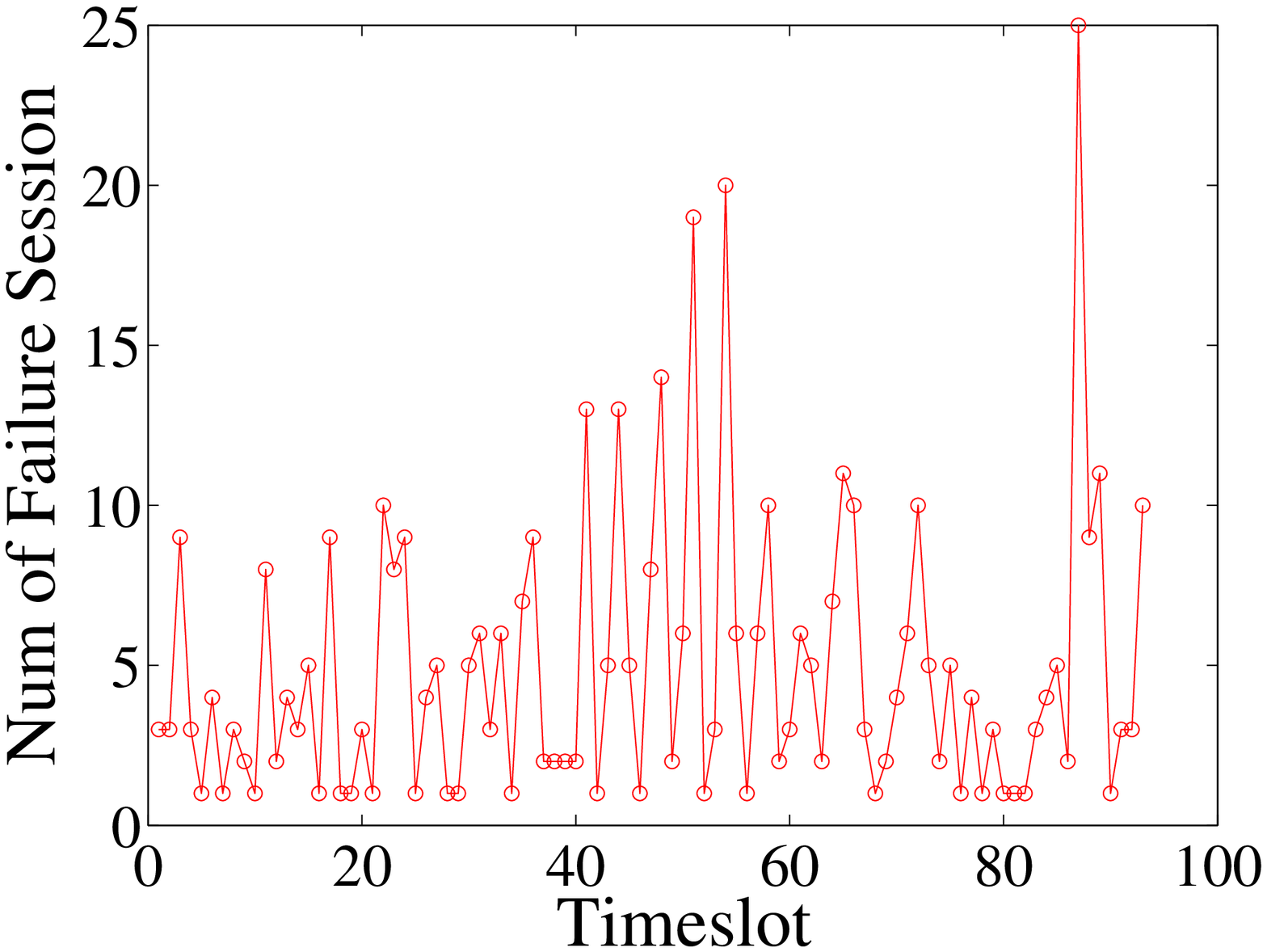}        
			  \caption{The number of failed sessions.}
		 \end{subfigure}
	\hfill
		\begin{subfigure}[b]{.45\linewidth}
	 	   \includegraphics[width=\linewidth]{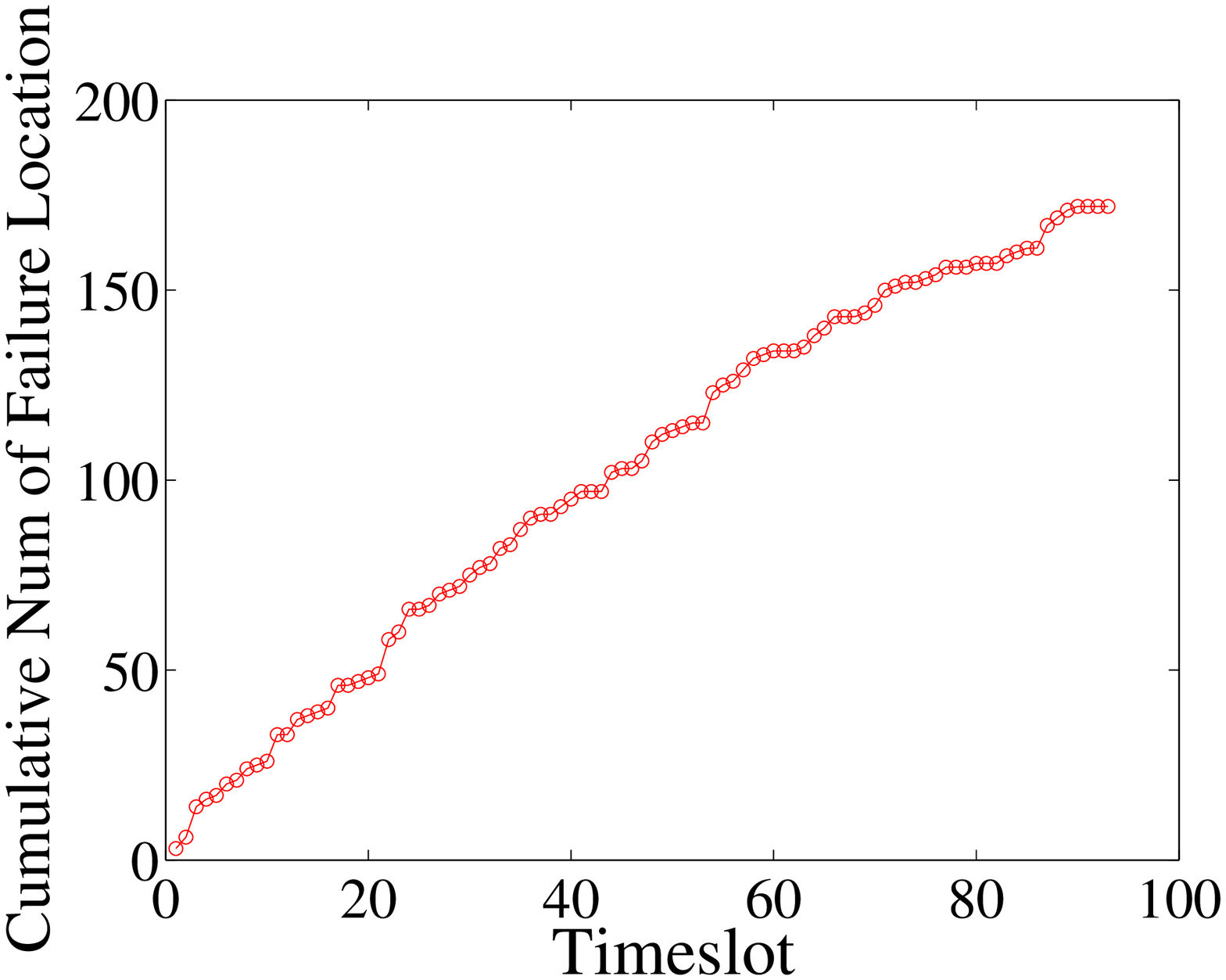}           
	 	  \caption{The cumulative number of failed users' locations.}
		\end{subfigure}
		\caption{Failure distribution over time. Each timeslot is six hours.} 
		\label{fig:failure_distribution}
	\end{minipage} 
	\vspace{-0.3cm}
	\hfill
     \begin{minipage}[t]{0.33\linewidth}
          \centering
               \includegraphics[width=\linewidth]{duration_distribution_cdf_all1.eps}        
			   \caption{Distribution of  webpage sessions' duration.} 
			   \label{fig:duration_distribution}     
	\end{minipage} 
	 \vspace{-0.3cm}
\end{figure*}

In our collected traces, we analyze the failure distribution in terms of the time and the locations. In Fig.~\ref{fig:failure_distribution}(a), we plot the number of content download failures recorded by our traces in each timeslot ($6$ hours). We observe that download failures may happen at different time of a day. Next, we plot the cumulative number of regions (city level) where users suffer from download failures over time in Fig.~\ref{fig:failure_distribution}(b). We observe that the number of cumulative locations where users have experienced network failures keeps increasing, indicating that the failure locations are geographically distributed, i.e., users located at many places may suffer from download issues. 

These observations indicate that the randomness of the content delivery failures makes the traditional replication/caching strategies based content delivery schemes no longer applicable.

Next, we study the design principles for our joint CDN and WebRTC-assisted webpage delivery framework, by measuring users' webpage accessing patterns.

\subsection{Webpage Viewing Patterns}

In our design, the peer delivery assistance is based on the implementation of WebRTC, which requires users to stay on a webpage so as to peer in content delivery. Thus, users' webpage viewing patterns have a significant impact on peers' resource availability.

\subsubsection{Extremely Short Session Duration}

As recorded in our collected traces, each webpage viewing event is defined as a ``session''. We first study the duration of the webpage viewing session, i.e., the time users spend on staying at a webpage. We have sampled $97,905$ users viewing two webpages (referred as Webpage A and Webpage B) of different types. In Fig.~\ref{fig:duration_distribution}, we plot the CDF of session durations of users viewing these two webpages on August, $2014$.

%

We have made the following observations: Compared with traditional P2P services and applications, webpage sessions have much shorter durations, i.e., over $60\%$ (resp. $90\%$) of session durations are shorter than $1$ minute ($10$ minutes). In a P2P content sharing service, how long peers stay in the system determines the level of peer resource \cite{wang2008stable}, since users are able to find more peering neighbors when peers stay longer. However, according to our observations of the extremely short webpage session durations, it is challenging for conventional P2P strategies to find enough peering resource to assist the web content delivery. Our design takes this short session characteristics into consideration and actively makes use of peers for content delivery assistance, by predicting when users will leave the webpage right after they open a webpage.

\subsubsection{Small Re-join Possibility}

\begin{figure}[t]
  \centering
     \begin{minipage}[t]{0.45\linewidth}
          \centering
			   \includegraphics[width=\linewidth]{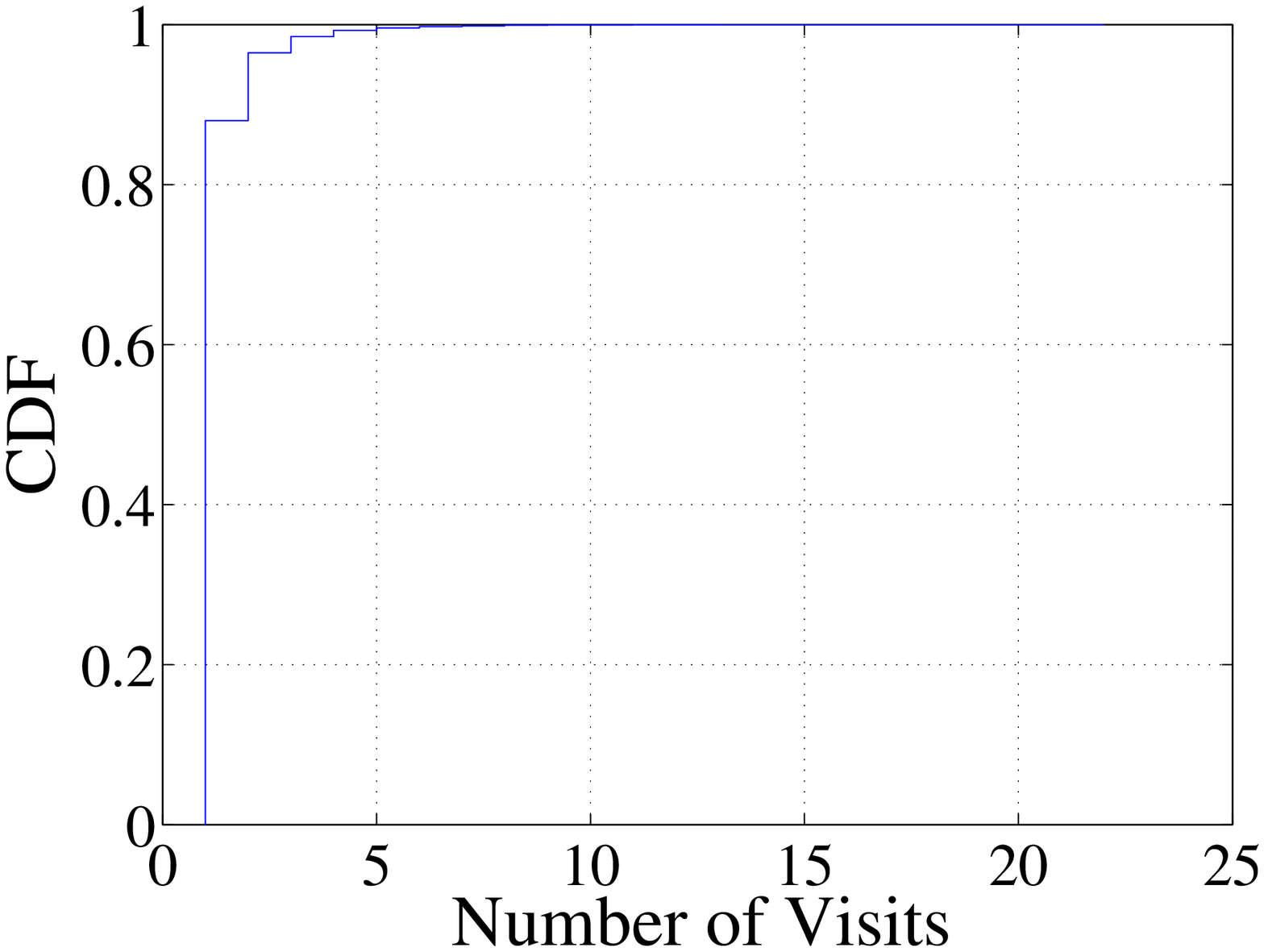}
				 \caption{CDF of number of visits to the same webpage.}
				 \label{fig:rejoin}
	\end{minipage}
	\hfill
     \begin{minipage}[t]{0.5\linewidth}
          \centering
		 \includegraphics[width=\linewidth]{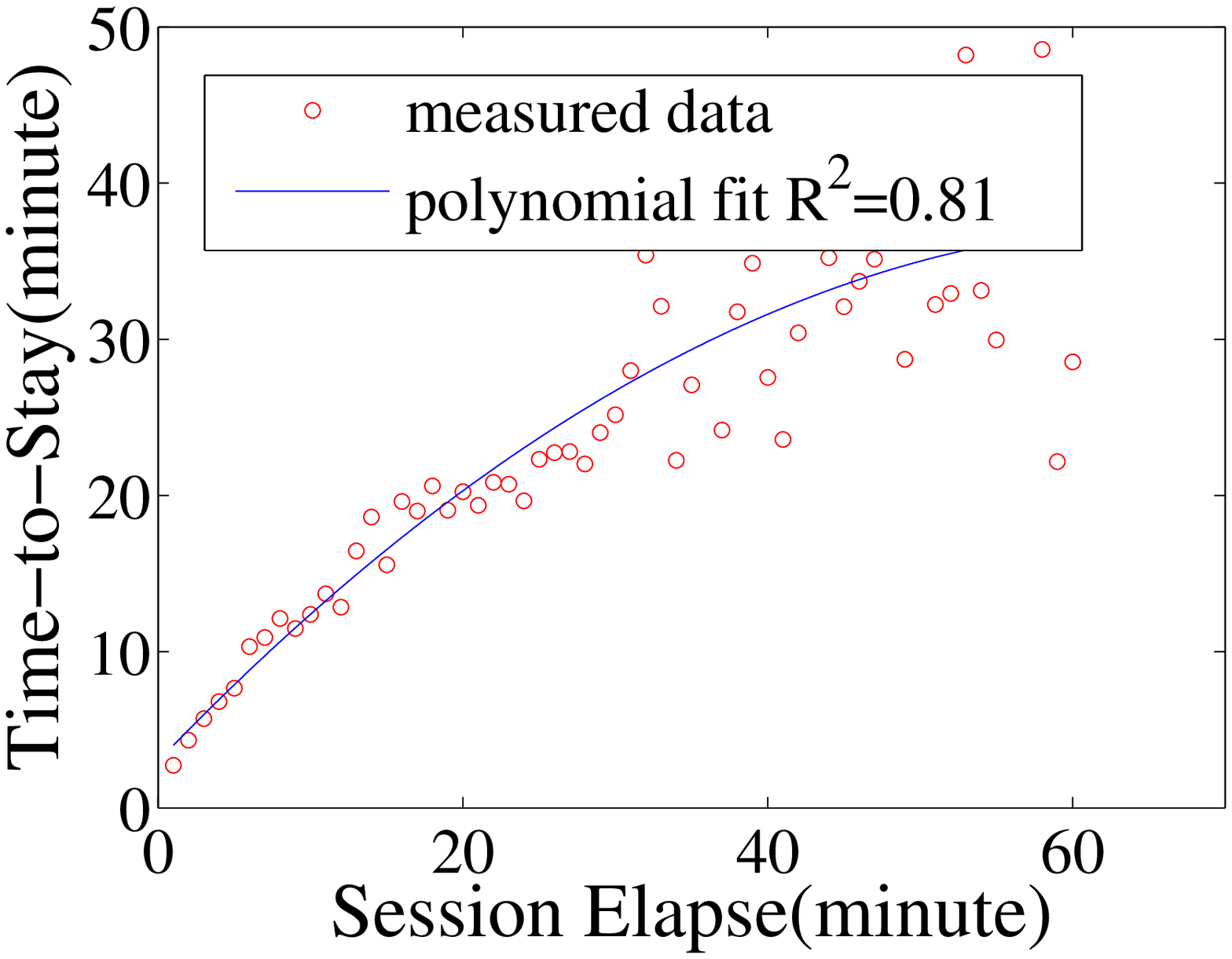}
		 \vspace{-0.5cm}
			 \caption{\emph{Time-to-stay} estimation for sessions \emph{elapse} in [0, 60].}
			 \label{fig:residual_duration}
     \end{minipage}
\end{figure}

In traditional P2P systems, peers re-joining the service may bring their cached data and serve other users. We also study the possibility that users re-join a webpage. As illustrated in Fig.~\ref{fig:rejoin}, the curve is the CDF of the number of visits from the same user to the same webpage in $14$ days. We observe that users hardly re-visit the webpages they have already viewed, e.g., less than $12\%$ of the users will return to the same webpage in $14$ days. The reason is that the two webpages in our experiments contain ``static'' content items, which do not change during our measurement period, and users do not re-visit these webpages to check out updates. As for webpages with more frequent updates, the re-join possibility may increase. 


This observation indicates that compared to traditional P2P sharing systems, whose clients may return to systems up to tens of times per day (e.g., $60$ times/day \cite{understand_p2p_churn}), the possibility of webpage re-visits is much smaller. This also challenges our CDN and peer-assisted web content delivery design, especially for webpages updated less frequently.

%

%

\subsubsection{High Correlation Between Time-to-Stay and Session Elapse}
\label{subsection:residual_lifetime}

As the short session durations may have a negative impact on the performance of our proposed CDN and peer-assisted web content delivery framework, we next study how to tell which peers may stay long in the system and which peers may leave the webpage soon. We take an in-depth look at the webpage session durations in our dataset. By randomly splitting session durations into two parts: the \emph{elapse} part, which is defined as the duration a user has already stayed on a webpage, and the \emph{time-to-stay} part, which is defined as the duration the user will keep staying on the webpage, we generate a set of (\emph{elapse}, \emph{time-to-stay}) pairs of each session. We plot the \emph{time-to-stay} versus the \emph{elapses} of $330$ samples from $30,000$ pairs (i.e., the session \emph{elapse} of $30,000$ pairs fall into $330$ elapse-time bins of one-minute length). Due to the space limitation, we do not show this figure. We observe that for the short sessions, the \emph{time-to-stay} is highly correlated with the \emph{elapse} of the sessions, indicating that a user tends to keep staying on a webpage if he/she has already spent longer time on the webpage. Furthermore, we dive into more details for the short samples (i.e., duration $< 60$ minutes), which accounts for $97\%$ of all the traces. As illustrated in Fig.~\ref{fig:residual_duration}, the polynomial model $y=-0.0076x^2 + 0.97x + 3.5$ fits well with these samples, further verifying the correlation between \emph{time-to-stay} and \emph{elapse}.

These observations and analysis indicate that we can accurately predict the \emph{time-to-stay} of a webpage session, and select ``stable'' peers in our joint CDN and peer-assisted delivery design.

\section{Path-Aware Peer-Assisted Web Content Delivery}
\label{section:algorithm}

In this section, we will analyze the peer selection problem in our CDN and peer-assisted framework. We formulate it as an integer programming problem, and design a heuristic algorithm to solve it.


\subsection{Framework}

We propose a path-aware peer-assisted web content delivery framework, where WebRTC peers are utilized to help other users fetch web content items from the web servers, to maximize the successful delivery ratio and the system throughput. We compare our content delivery framework with the conventional web content delivery paradigm in Fig.~\ref{fig:data_relay_framework}. Fig.~\ref{fig:data_relay_framework}(a) presents the conventional content delivery paradigm, where users directly download content items contained in a webpage from the corresponding CDN servers. Fig.~\ref{fig:data_relay_framework}(b) presents our peer-assisted delivery paradigm: In the case of delivery failures (denoted as a red cross), a user (e.g., Alex) will try to receive web content items from other WebRTC peers (e.g., Bob) by a recovered path (denoted as green dotted lines).



\subsection{Problem Formulation}
The main idea behind our design is to strategically assign the WebRTC peers, which have available peering resource (e.g., uplink bandwidth capacity), to users who encounter delivery failures. As such, generating a candidate peer list through carefully peer selection for each failed user is critical in our design. Before diving into more details, we present several definitions used in our formulation.

%

\subsubsection{P2P Connectivity and Bandwidth Matrix}

First, we define a peer-to-peer bandwidth matrix $\textbf{B}^{M \times N}$ and each entry $b_{ij}(t)$ represents the estimated upload capacity from a WebRTC peer $j$ to user $i$ at time $t$.




\subsubsection{Peer Selection Strategy}

Next, we define our peer selective strategy as a peer selection matrix $\textbf{P}^{N \times M}$ and each entry is a binary variable, i.e., $p_{ij}(t) = 1$ indicates we choose peer $i$ as the relay peer for peer $j$ at time $t$; otherwise, $p_{ij}(t) = 0$. In our design, the peer selection is generated over time, i.e., the peer selection matrix $\textbf{P}$ changes over time according to the estimated P2P bandwidth. We assume that a user only utilizes one relay peer when downloading a single content. However, our design can also be extended to use multiple relay peers to fetch a web content.



\subsubsection{Optimization Formulation}

We formulate the peer selection problem as an optimization problem, as follows.
\begin{equation} 
  \max_{p_{ij}(t)} \sum_{i = 1}^n \sum_{j = 1}^m p_{ij}(t) b_{ji}(t) 
\end{equation}
subject to,
\begin{equation}\label{eq:at_least_one}
  \sum_{i = 1}^n p_{ij}(t) \geq 1 \; \; \; \; \; \; \; \; \; \; \; \; \; \; \; \; \; \; \; \;\forall j \in \{1,\dots, m\} 
\end{equation}
\begin{equation}\label{eq:not_exceed_bandwidth}
  \sum_{j = 1}^m p_{ij}(t) b_{ji}(t) \leq bw_{i}(t) \; \; \; \; \forall i \in \{1,\dots, n\}
\end{equation}
The rationale of the optimization is to find a ``match'' between relay peers and the users encountering delivery issues, to maximize the successful relay ratio and the throughput of delivering the failed content items. Constraint (\ref{eq:at_least_one}) guarantees that there is at least one peer available for the requesting user. Constraint (\ref{eq:not_exceed_bandwidth}) guarantees that the relay peer can not serve many requests exceeding its own uplink capacity.

\subsection{Path-Aware Peer Selection Strategy}

Our heuristic and distributed algorithm works as follows: (1) Based on various peer selection factors, we generate a peer list for a requesting user; (2) The requesting user then actively tries these candidate peers to download the failed web content. We next elaborate these two steps, respectively.

\subsubsection{Factors for Peer Selection}
First, since the real-time measurement of network state is too expensive and the geographic location and the ISP of the peer provide valuable implications for the network state~\cite{hu2015guyot}, in our design, we select relay peers that are within the same location and Internet Service Providers (ISP) with the requesting peers in order to achieve a better network performance. In particular, the same ISP also reduce the cross ISP cost~\cite{bittorrent}.
 
Moreover, due to the extremely short duration of web viewing sessions, it is of paramount importance to guarantee that the relay peer will still be online when it is chosen to serve the requesting peer. Thus, we have to estimate how long a peer will be online. In our measurement study in Section~\ref{subsection:residual_lifetime}, we observe the fact that the \emph{time-to-stay} of a session is highly correlated with the \emph{elapse} of the session. Based on this observation, we prioritize candidate relay peers according to the duration they have already been viewing a webpage. As such, the selected relay peers are more likely to be online during the content delivery phase.

Besides, according to our measurement studies, content delivery failures show randomness with respect to the time and the locations. In order to increase the diversity of potential content delivery paths, we (1) select relay peers who are at different locations to overcome the regional network failure; (2) choose relay peers with diverse ISPs such that relay peers may have an improved connectivity with the original content server; (3) select the peers with various load to achieve load balance among all relay peers in the system.

\subsubsection{Relay Peer List Generation} 

Based on the factors discussed above, when there is a relay request from peer $p_s$, two steps are involved to generate the relay peer list.
\begin{itemize}
	\item The requesting peer $p_s$ chooses relay peer $p_r$ from the online peer set, which is obtained from the coordinate server;
	\item Rank the list based on the workload and the \emph{time-to-stay} of the candidate relay peers.
\end{itemize}


The peer selection algorithm carried out by the coordinate server is summarized in Algorithm \ref{algorithm:peer_selection}. Given the input parameter $\zeta$, which is the length of relay peer list, we generate the list consisting of two peer sets: the peers selected carefully ($R_I$) and the peers selected randomly ($R_{II}$). $\alpha$ is the ratio of each set and it is adjustable based on the specific network environment. Note that $R_I$ consists of peers located in the same city and served by the same ISP with the requesting peer (line~\ref{line:same_property}), to achieve a better network performance, and $R_{II}$ consists of randomly selected peers (line~\ref{line:random_property}), to increase the diversity of potential content delivery paths. We filter those relay peers which experienced some content fetching failures or a higher workload (i.e., how many peers it has been already relaying for) exceeding the threshold $\gamma$ in the two sets ($R_I, R_{II}$) (line~\ref{line:filter_failure}, line~\ref{line:filter_overload}). Furthermore, in order to select the peers not only have more spared bandwidth and will stay online longer, we sort the two parts in the descending order of the \emph{time-to-stay} ${\tau}_i$ (line~\ref{line:sort}).

\subsubsection{Content Relaying with Relay Peers}
The first peer in the candidate relay list acts as the \emph{primary} peer, and the remaining peers act as backups. After obtaining the relay peer list, the content relaying works as follows: The requesting peer $p_s$ first asks the primary peer to fetch the failed content, then it tries other candidates in the candidate list until the content is finally fetched. 


%
%
%
%

\setlength{\textfloatsep}{0.1cm}
\begin{algorithm}[t]
  \caption{Relay Peer List Generation.}\label{algorithm:peer_selection}
  \begin{algorithmic}[1]
	\Procedure{relay-peer-selection}{$\alpha$, $\gamma$, $\zeta$}
	\For {$i = 0$ to $\zeta * \alpha$} 
	\State \textbf{select} $p_r$ which has the same location and ISP with $p_s$, from online peer set \label{line:same_property}
	\State $R_I \gets R_I \cup p_r$
	\EndFor 
	\For {$i =  (\zeta * \alpha + 1)$ to $\zeta$}
	\State \textbf{select} $p_r$ from online peer set randomly \label{line:random_property}
	\State $R_{II} \gets R_{II} \cup p_r$
	\EndFor 
	\State \textbf{filter} peers which experienced some content fetching failure in $R_I,R_{II}$, respectively \label{line:filter_failure}
	\State \textbf{filter} peers whose workload exceed the threshold $\gamma$ in $R_I,R_{II}$, respectively \label{line:filter_overload}
	\State \textbf{sort} $R_I,R_{II}$ in descending order of ${\tau}_i$ \label{line:sort}
	\State $R \gets R_I \cup R_{II}$
  \Return $R$ 
  \EndProcedure
  \end{algorithmic}
\end{algorithm}

\section{Performance Evaluation}
\label{section:experiment}

We verify the effectiveness and performance of our design with simulation experiments driven by the traces presented in Section~\ref{section:measurement}.

\subsection{Experiment Setup}
To study the details in its performance under extreme scenarios, we carry out simulation experiments with different settings. In our experiments, we simulate $5,000$ peers that are distributed in five cities in China. Based on the geographic of the cities, i.e., the longitude and latitude information, we calculate the physical distance between each peer pair. The latency between peer pairs is estimated using the correlation between distance and latency \cite{ip_geolocation}. To simulate the real Internet environment, the peers are configured heterogeneously in terms of uplink and downlink capacities, following the statistics in \cite{p2pstream}. The ISP of each user is randomly assigned to $3$ ISPs. According to the survey in \cite{webpage_size}, the average web page is $1,600$ KB consisting of 112 objects. Here, we analyze the content size ranging from $500$ KB to $16,000$ KB due to the rapid growth of the web page size.
%
%

\textbf{User behaviors:} We simulate the arrival patterns as a poisson process with the $\lambda=30$, and the session duration with the Pareto distribution observed in Section ~\ref{section:measurement}. Relaying peers who are potential to help the failed peer fetch content are selected according to Algorithm \ref{algorithm:peer_selection}. If not otherwise specified, $\alpha=0.2$, $\gamma=0.8$, $\zeta=10$~\cite{mesh_p2p_streaming}.
 
\begin{figure}[t]
          \centering
               \includegraphics[width=.6\linewidth]{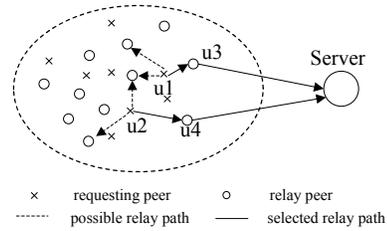}
     \caption{Illustration of in-network failures.}
     \label{fig:error_case}
\end{figure}

\textbf{Content delivery failures:} In our experiments, we assume that users download web content items from regional CDN servers, in which users are redirected to a regional peering server to download the content items according to their locations and ISPs. We simulate in-network failures as follows. An in-network failure \cite{wan_service} is the case that the network of a user's region is encountering some issues, and a fraction of users in this region can not connect to the server. An example of such failure is illustrated in Fig.~\ref{fig:error_case}: user $u_1$ and user $u_2$ can not download content from the server and connect to each other; however, the other users/peers (e.g., $u_3$, $u_4$), are able to connect to both $u_1$, $u_2$ and the server. In our experiments, the in-network failure ratio is $60\%$ if not otherwise specified.



\begin{figure*}[t]
     \begin{minipage}[t]{.33\linewidth}
          \centering
               \includegraphics[width=\linewidth]{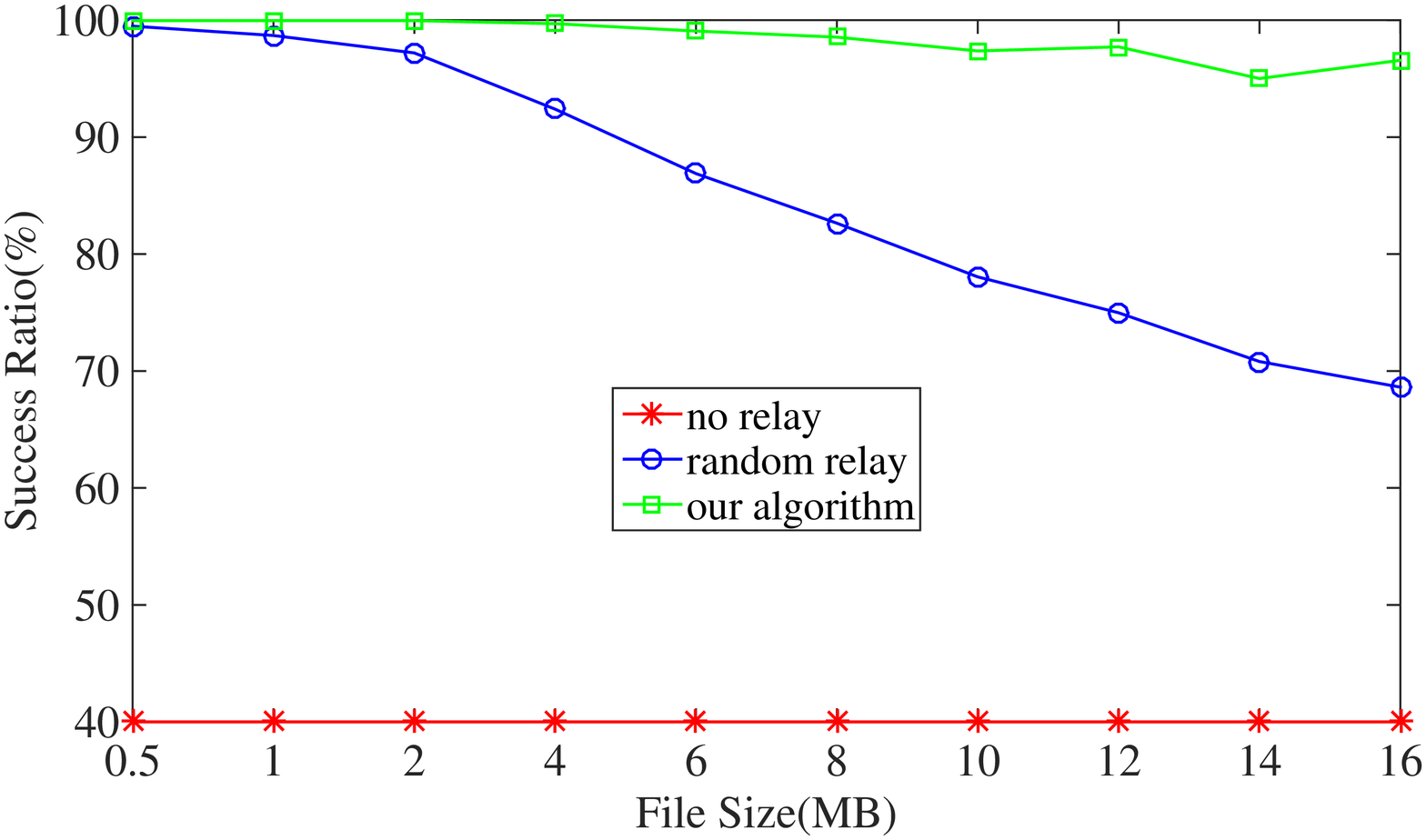} 
			   \raggedright        
		\scriptsize (a) Average successful relay ratio vs. file size.      
	\end{minipage}      
	\hfill
     \begin{minipage}[t]{.33\linewidth}
          \centering      
         \includegraphics[width=\linewidth]{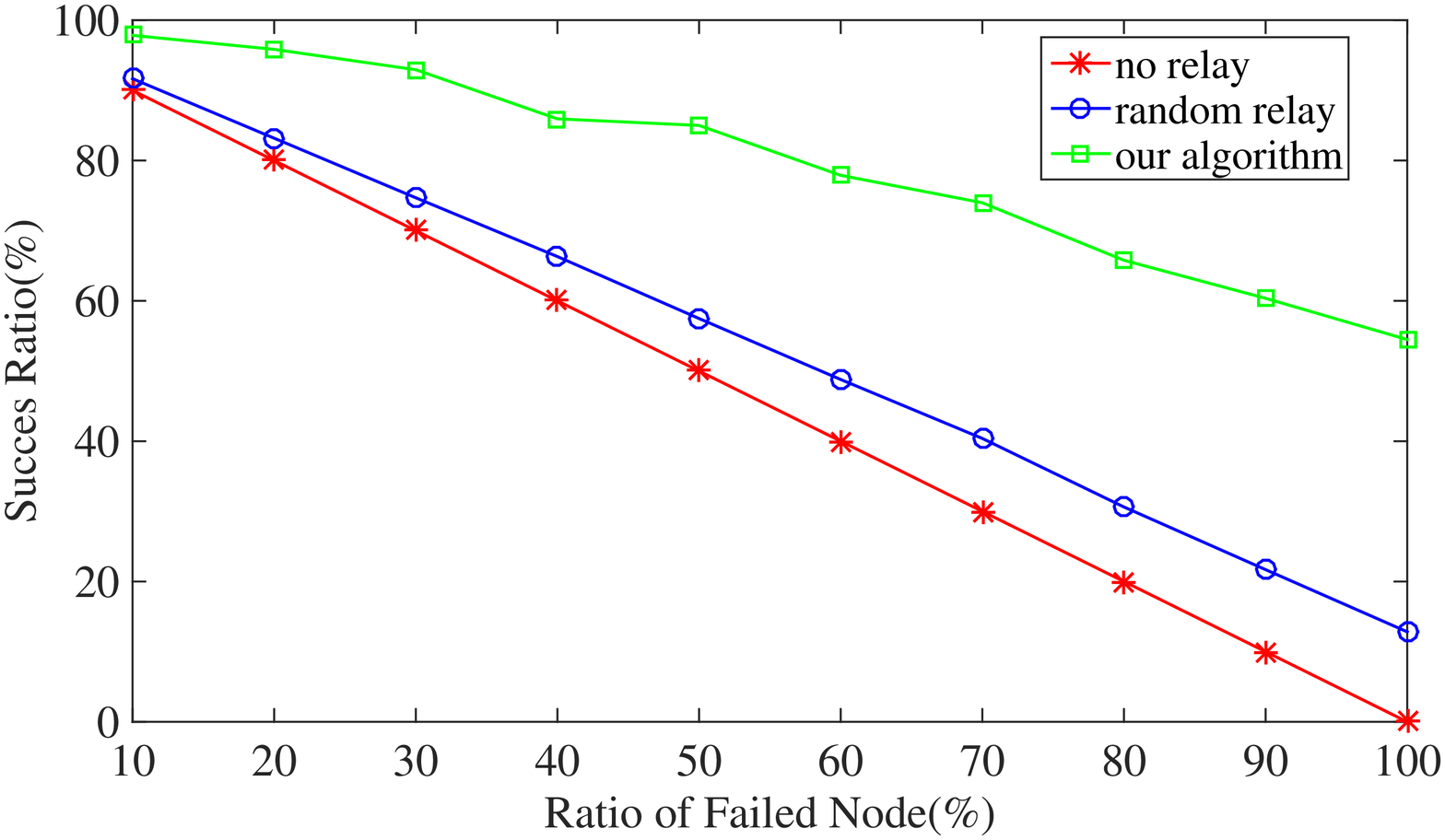}    
        \raggedright
		\scriptsize (b) Successful ratio with the primary peer vs. ratio of failed nodes.
     \end{minipage}
	\hfill
     \begin{minipage}[t]{0.32\linewidth}
          \centering
               \includegraphics[width=\linewidth]{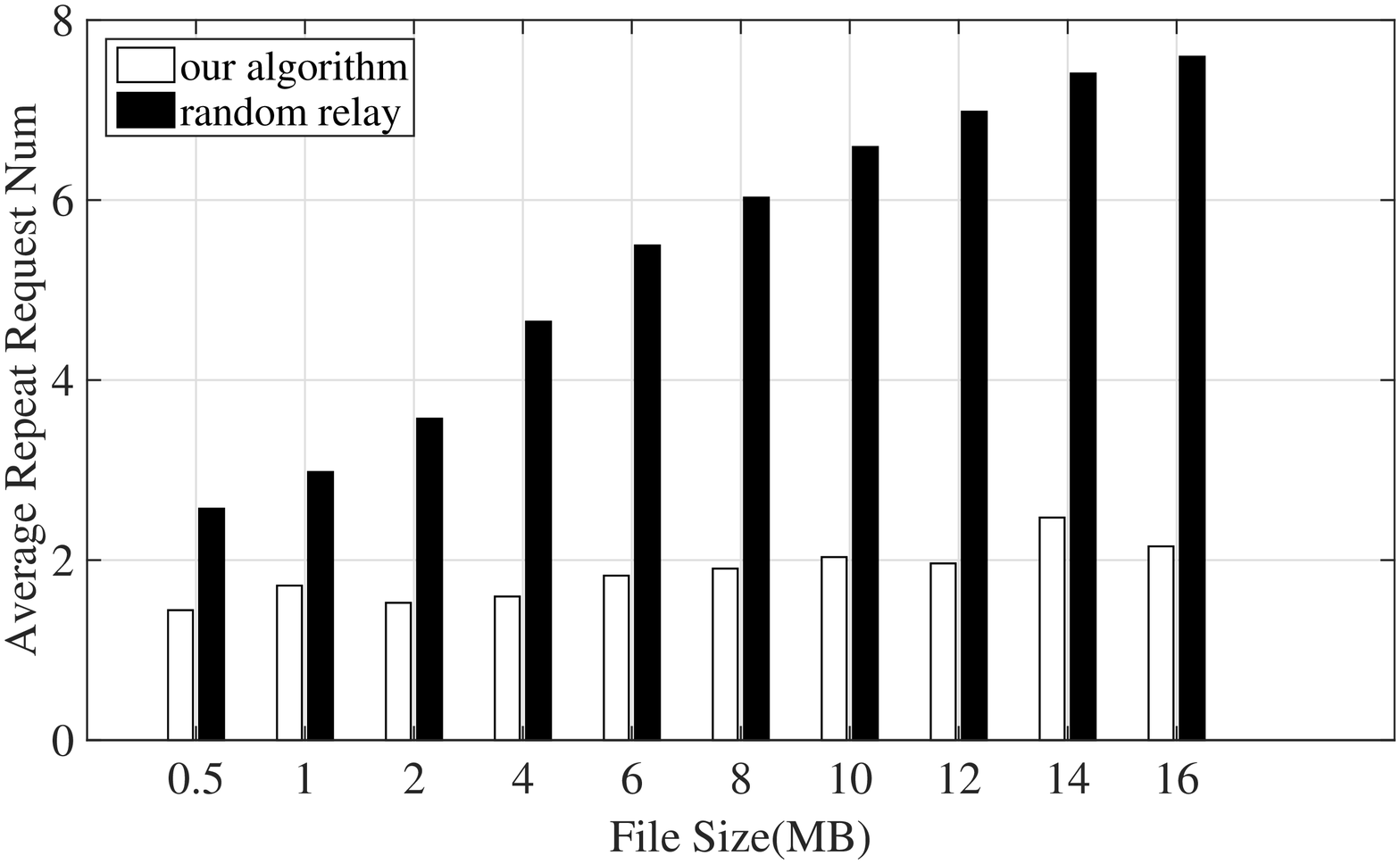}  
			   \raggedright         
			   \scriptsize (c) Average repeated request number of successful relays vs. file size.  
			   		  
	\end{minipage} 
     \caption{The performance comparison among different algorithms when in-network failures occur.}
     \label{fig:peer_error_evaluation}
	 	 \vspace{-0.4cm}
\end{figure*}


\textbf{Baselines:} We compare our design with two schemes, i.e., (1) no relay: users directly download content items from CDN servers; (2) random peer selection: users request randomly selected peers to download failed content items; (3) our design: users download failed content items from the peers selected with our strategies. Next, we present the experiment results.

\subsection{Performance under In-Network Failures}

First, we study the successful relay ratio, which is defined as the ratio of requests that can be finally served by CDN servers or peers. As illustrated in Fig.~\ref{fig:peer_error_evaluation}(a), the curves represent the successful relay ratio of different strategies versus the web content size. We observe that our design outperforms the random peer selection strategy, and the no-relay strategy. In particular, when the content size is about $16$ MB, our design improves the successful content download ratio by over $25\%$ against the random relay selection approach. 

Next, we investigate the impact of the in-network failure ratio on the successful relay ratio with the primary peer, i.e., the first relay peer in the candidate relay peer list. As illustrate in Fig.~\ref{fig:peer_error_evaluation}(b), the curves represent the successful relay ratio versus the in-network failure ratio. We observe that with the increase of in-network failure ratio, the successful relay ratio decreases. In particular, compared with our strategy, the successful relay ratio of random and no-relay strategy decrease much faster. When the in-network failure ratio reaches $100\%$, i.e., all users in a particular region can not connect to the CDN server, our design still achieves a final successful download ratio of $60\%$. 




We further investigate the average repeated request number in the successful relay session, which is defined as the number of repeated requests from requesting users to the candidate relay peers until it successfully downloads the content. In Fig.~\ref{fig:peer_error_evaluation}(c), the bars represent the number of requests tried by requesting users versus the file size of the web content items. We observe that compared with the random peer selection, users try a much smaller number of relay requests to successfully download a failed content, and the performance gap is more remarkable when the file size increases. In particular, when the file size is $16$ MB, it only takes about $2$ tries for a user to download a content from the relay peers selected with our strategy, whereas it takes more than $7$ tries with the random peer selection approach.

%


\section{Related Work}
\label{section:relatedwork}

In this section, we survey related works on WebRTC-powered content delivery.


WebRTC allows direct browser-to-browser communication, leading to growing peer-assisted web applications. To date, there are over $1$ billion endpoints in use supporting WebRTC, and it is anticipated that $4.7$ billion devices will support WebRTC by 2018 \cite{alexandru2014impact}. Some systems have been developed to utilize WebRTC browsers to assist in content delivery. Zhou et al.~\cite{zhou2012webcloud} developed WebCloud, which decentralized web content exchanging by utilizing users' browsers to deliver content; however, it required that redirector proxies had been deployed within each ISP region. Zhang et al.~\cite{zhang2013maygh} proposed Maygh, a system that automatically recruited web visitors to help content servers, to reduce the cost for operating a web site. 


The limitation of the previous works on WebRTC is that they mainly focused on alleviating the original content servers based on the P2P philosophy, in which peers share bandwidth with each other. However, our work focuses on seeking a solution to address the network failure problem~\cite{hu2015path}.

\section{Conclusion}
\label{section:conclusion}


In this paper, we propose a joint CDN and peer-assisted web content delivery framework to address the content delivery failure problem for today's WebRTC-powered peers. Different from traditional peer-assisted approaches that mainly focus on alleviating the CDN servers' bandwidth load, our contribution in this paper lies in that we first study to use a browser-based peer-assisted scheme to resolve content delivery failure problem. Based on large-scale measurement studies on how users access and view webpages, we not only show the challenges in our design that can not be directly solved by conventional P2P strategies (e.g., peers stay on a webpage extremely short), but also learn webpage viewing patterns and design principles. In particular, we formulate the peer selection problem as an optimization problem, and design a heuristic algorithm based on the measurement insights to solve it. Our simulation experiments demonstrate the effectiveness of our design: compared with non-peer-assisted strategy and random peer selection strategy, our design significantly improves the successful delivery ratio under network failures.




\bibliographystyle{IEEEtran}
\bibliography{mylib_short}

\end{document}